\RequirePackage{ifpdf}
\documentclass[12pt,letterpaper]{article}
\usepackage{jheppub}
\usepackage{epsfig}
\usepackage[latin1]{inputenc}
\usepackage{bbm,amsfonts}
\usepackage{graphicx}
\usepackage{amssymb,amsmath}
\usepackage{fancybox}
\usepackage{verbatim}

\author[a]{Marco S. Bianchi,}
\author[b]{ Matias Leoni}
\preprint{QMUL-PH-16-22}
\affiliation[a]{Center for Research in String Theory - School of Physics and Astronomy Queen Mary University of London, Mile End Road, London E1 4NS, UK}
\affiliation[b]{Physics Department, FCEyN-UBA \& IFIBA-CONICET\ 
  Ciudad Universitaria, Pabell\'on I, 1428, Buenos Aires, Argentina }

\emailAdd{m.s.bianchi@qmul.ac.uk}  
\emailAdd{leoni@df.uba.ar}

\title{A $QQ\to QQ$ planar doublebox in canonical form}
\abstract{We consider a planar doublebox with four massive external momenta and two massive internal propagators. We derive the system of differential equations for the relevant master integrals, cast it in canonical form, write it as a $d\log$ form and solve it in terms of iterated integrals up to depth four.}

\csname @addtoreset\endcsname{equation}{section}

\def\bseq{\begin{subequation}}  % = 1a 1b
\def\eseq{\end{subequation}}
\def\bsea{\begin{subeqnarray}}  % = 1.1a 1.1b
\def\esea{\end{subeqnarray}}
\newcommand{\beq}{\begin{equation}}
\newcommand{\bea}{\begin{eqnarray}}
\newcommand{\eea}{\end{eqnarray}}
\newcommand{\eeq}{\end{equation}}

\def\beq{\begin{equation}}
\def\eeq{\end{equation}}
\def\bea{\begin{eqnarray}}
\def\eea{\end{eqnarray}}

\begin{document}
\maketitle
\allowdisplaybreaks

\section{Introduction}

Differential equations are a powerful tool for solving master integrals \cite{Kotikov:1990kg,Kotikov:1991pm,Bern:1993kr,Remiddi:1997ny,Gehrmann:1999as,Gehrmann:2000zt,Gehrmann:2001ck}.
This method received considerable boost by the observation that the system of differential equations can be often put in a canonical form, by a suitable choice of master integrals exhibiting uniform transcendentality \cite{Henn:2013pwa} (see also \cite{Kotikov:2010gf} and the review \cite{Henn:2014qga}). In such a form the solution of the differential equations becomes much simpler and lands on Chen iterated integrals \cite{Chen:1977oja} which in many cases evaluate to harmonic \cite{Remiddi:1999ew} and Goncharov polylogarithms \cite{Goncharov:1998kja}.
This technique has proven efficient and has been applied in a number of contexts \cite{Henn:2013fah,Henn:2013woa,Caola:2014lpa,Argeri:2014qva,Gehrmann:2014bfa,
vonManteuffel:2014mva,Dulat:2014mda,Gehrmann:2015ora,Henn:2016men,Bonciani:2016qxi,
Henn:2016kjz,Lee:2016ixa}.

\begin{figure}
\begin{center}
\includegraphics[scale=0.7]{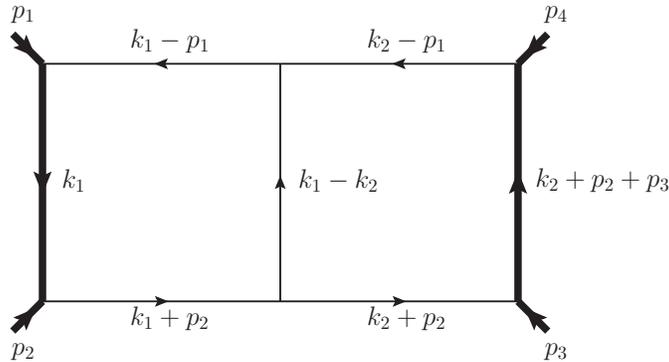}
\caption{Momentum labels of the propagators. The figure does not include the irreducible numerators $P_4$ and $P_6$. The thick internal lines are the massive propagators.}
\end{center}
\label{fig:momentum}
\end{figure}

In this short note, we apply this method to a planar doublebox topology with four external massive momenta $p_1^2=p_2^2=p_3^2=p_4^2=-m^2$ and two internal massive propagators, all with the same mass $m$. The integral is depicted in Figure \ref{fig:momentum}.
More explicitly, we consider the class of two-loop integrals given by the following propagators
\begin{equation}
g_{a_1,\dots,a_9}=
\frac{e^{2\epsilon\gamma_{E}}}{(\pi^{d/2})^2}\, \int d^{d} k_1\, d^{d} k_2\, \frac{P_4^{-a_4}P_6^{-a_6}}{P_1^{a_1}\, P_2^{a_2}\, P_3^{a_3}\, P_5^{a_5}\, P_7^{a_7}\, P_8^{a_8}\, P_9^{a_9}}
\end{equation}
where $d=4-2\epsilon$ and
\begin{equation}
\begin{array}{ccccc}
P_1 = (k_1-p_1)^2 & \quad & P_2= k_1^2+m^2 & \quad & P_3= (k_1+p_2)^2 \\ 
P_4= (k_1+p_2+p_3)^2+m^2 & \quad & P_5 = (k_2-p_1)^2 & \quad & P_6= k_2^2+m^2\\
P_7= (k_2+p_2)^2 & \quad & P_8= (k_2+p_2+p_3)^2+m^2 & \quad & P_9=(k_1-k_2)^2
\end{array}
\end{equation}
where we work in the Euclidean and we assume that $a_4\leq0$ and $a_6\leq0$, namely these correspond to two irreducible numerators of the doublebox.
This kind of integrals arises (among others) in $2\to 2$ scattering of massive particles which can exchange massless non-abelian gauge bosons, such as massive quarks in QCD.
Such a topology can also be considered as an extension of that studied in \cite{Henn:2013pwa} (albeit with a different labelling of propagators and momenta),  where the roles of two numerators/propagators have been exchanged.
Indeed, several master integrals relevant in this model were considered in literature before, mostly in the context of Bhabha scattering, using again the differential equations approach or evaluation via Mellin-Barnes representation \cite{Smirnov:2001cm,Bonciani:2003cj,Bonciani:2004gi,Heinrich:2004iq,Czakon:2004wm,Czakon:2006pa}.
On the contrary, to the best of our knowledge, the seven propagators topologies presented in this note are new in literature. 

The problem has three quadratic invariants, $s\equiv (p_1+p_2)^2$, $t\equiv (p_2+p_3)^2$ and $m^2$, giving rise to two adimensional ratios.
In the following it proves convenient to express the integrals in terms of the dimensionless variables
\begin{equation}\label{eq:changeofvars}
\frac{s}{m^2} = \frac{(1-x)^2}{x} \qquad\qquad \frac{t}{m^2} = \frac{(1-y)^2}{y}
\end{equation}
which are commonly used for massive particles scattering and rationalize typical square roots arising in this context.

In the rest of the article we provide the details regarding the system of differential equations and the basis of master integrals of uniform transcendentality that allows to write it in canonical form. Then we cast the latter in $d\log$ form depending on letters chosen from an alphabet of 16.
Finally we sketch the solution of the differential equations in terms of iterated integrals, most of which are expressed in terms of Goncharov and harmonic polylogarithms (whose definition we review in Appendix \ref{app:polylogs}), and the procedure to fix the integration constants. The results up to depth four are collected in electronic format in the ancillary file \texttt{results.m}.

\section{Basis of master integrals}

\begin{figure}
\includegraphics[width=\textwidth]{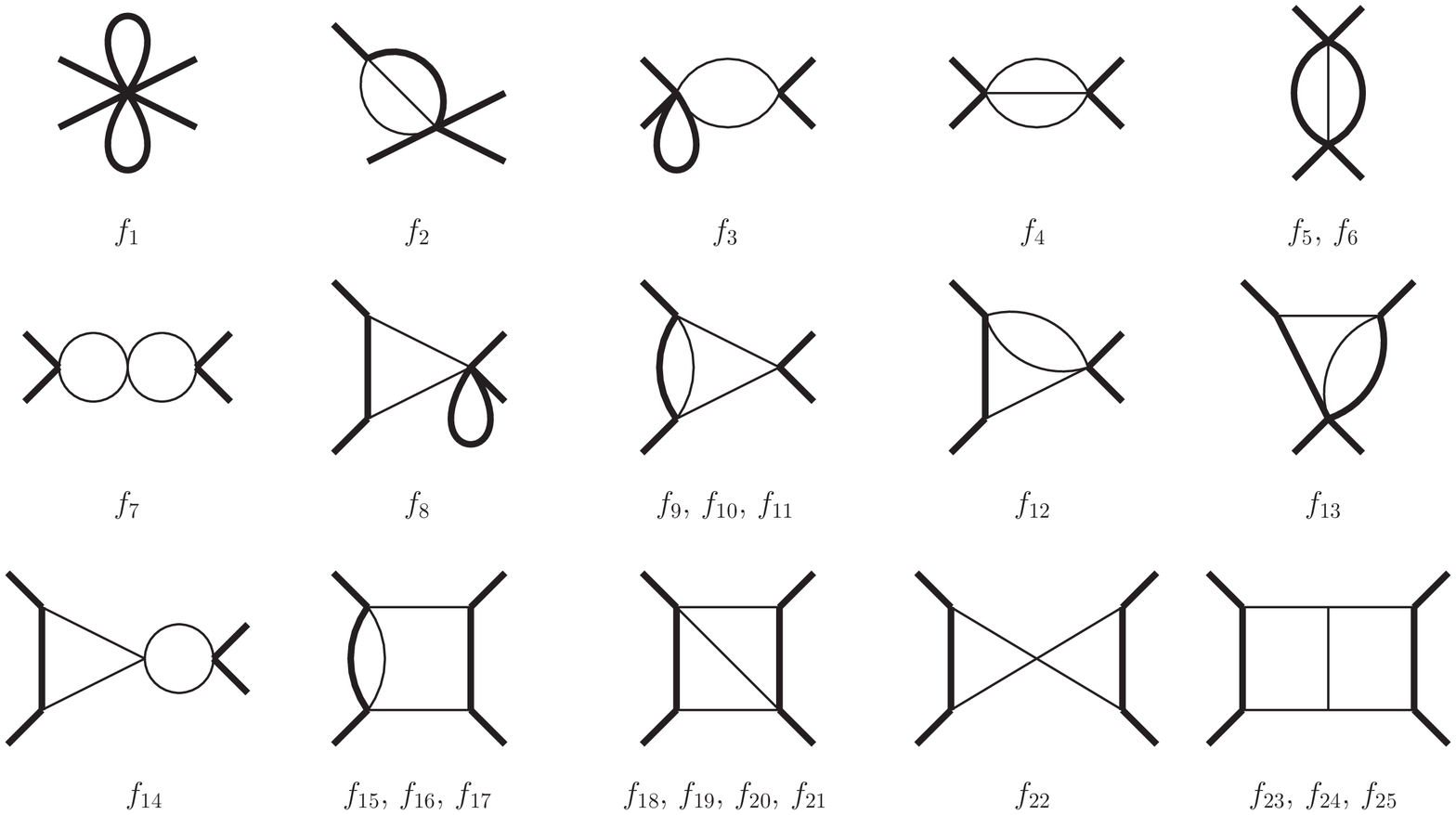}
\caption{Master integral topologies.}
\label{fig:MI}
\end{figure}

The system possesses 25 master integrals, which can be put in canonical form as follows:
\begin{align}
f_1 &= \epsilon ^2\, g_{0,2,0,0,0,0,0,2,0} \\
f_2 &= m^2\, \frac{\epsilon ^2 (4 \epsilon +1) g_{0,1,0,0,2,0,0,0,2}}{\epsilon +1} \\
f_3 &= s\, \epsilon ^2\, g_{0,2,0,0,2,0,1,0,0} \\
f_4 &= s\, \epsilon ^2\, g_{2,0,0,0,0,0,2,0,1} \\
f_5 &= \frac{1}{2} \sqrt{t} \sqrt{4 m^2+t}\, \epsilon ^2\, \left(2 g_{0,2,0,0,0,0,0,1,2}+g_{0,2,0,0,0,0,0,2,1}\right) \\
f_6 &= t\, \epsilon ^2\, g_{0,2,0,0,0,0,0,2,1} \\
f_7 &= s^2\, \epsilon ^2\, g_{2,0,1,0,2,0,1,0,0} \\
f_8 &= \sqrt{s} \sqrt{4 m^2+s}\, \epsilon ^3\, g_{0,2,0,0,1,0,1,1,0} \\
f_9 &= \sqrt{s} \sqrt{4 m^2+s}\, \epsilon ^3\, g_{0,2,0,0,1,0,1,0,1} \\
f_{10} &= m^2\, \sqrt{s} \sqrt{4 m^2+s}\, \epsilon ^2\, g_{0,3,0,0,1,0,1,0,1} \\
f_{11} &= s\, \epsilon ^2 \left(\frac{3}{2}\, \epsilon\, g_{0,2,0,0,1,0,1,0,1}+m^2\, g_{0,2,0,0,2,0,1,0,1}-m^2\, g_{0,3,0,0,1,0,1,0,1}\right) \\
f_{12} &= 2\, \sqrt{s} \sqrt{4 m^2+s}\, \epsilon ^3\, g_{0,1,1,0,2,0,0,0,1} \\
f_{13} &= 2\, \sqrt{t} \sqrt{4 m^2+t}\, \epsilon ^3\, g_{0,2,0,0,1,0,0,1,1} \\
f_{14} &= s^{3/2}\, \sqrt{4 m^2+s}\, \epsilon ^3\, g_{1,1,1,0,2,0,1,0,0} \\
f_{15} &= \sqrt{s} \sqrt{4 m^6+s (t+m^2)^2}\, \epsilon ^3\, g_{0,2,0,0,1,0,1,1,1} \\
f_{16} &= \sqrt{s} \sqrt{4 m^2+s}\, \epsilon ^3\, \left(g_{0,2,0,0,1,-1,1,1,1}-m^2\, g_{0,2,0,0,1,0,1,1,1}\right) \\
f_{17} &= s\, \sqrt{t} \sqrt{4 m^2+t}\, \epsilon ^2\, \left(m^2\, g_{0,3,0,0,1,0,1,1,1}-\epsilon\,  g_{0,2,0,0,1,0,1,1,1}\right) \\
f_{18} &= 4\, \sqrt{s+t} \sqrt{4 m^2+s+t}\, \epsilon ^4\, g_{1,1,0,0,0,0,1,1,1} \\
f_{19} &= 2\, s\, \sqrt{t} \sqrt{4 m^2+t}\, \epsilon ^3\, g_{1,1,0,0,0,0,1,1,2} \\
f_{20} &= 2\, m^2\, \sqrt{s} \sqrt{4 m^2+s}\, \epsilon ^3\, g_{1,2,0,0,0,0,1,1,1} \\
f_{21} &= \frac{1}{2} s\, \epsilon ^2\, \big[ 2\, m^2 \left(m^2\, g_{1,2,0,0,0,0,1,2,1}-2 \epsilon\,  g_{1,2,0,0,0,0,1,1,1}\right)-\left(2 m^2+t\right) \epsilon\,  g_{1,1,0,0,0,0,1,1,2}\big] \\
f_{22} &= s \left(4 m^2+s\right)\, \epsilon ^4\, g_{1,1,1,0,1,0,1,1,0} \\
f_{23} &= s^2\, \sqrt{t} \sqrt{4 m^2+t}\, \epsilon ^4\, g_{1,1,1,0,1,0,1,1,1} \\
f_{24} &= s^{3/2}\, \sqrt{4 m^2+s}\, \epsilon ^4\, g_{1,1,1,-1,1,0,1,1,1} \\
f_{25} &= \frac{1}{2} s\, \epsilon ^2\, \Big[-4 m^2\, g_{0,3,0,0,1,0,1,0,1}+\frac{g_{2,0,0,0,0,0,2,0,1}}{1-2 \epsilon }+2 \epsilon  \big(-4 g_{0,1,1,0,2,0,0,0,1}+\\ & 
+2 g_{0,2,0,0,1,-1,1,1,1}+g_{0,2,0,0,1,0,1,0,1}-g_{0,2,0,0,1,0,1,1,0}-2 m^2\, g_{0,2,0,0,1,0,1,1,1}+ \nonumber\\& +s\, g_{1,1,1,0,2,0,1,0,0}+2 m^2\, g_{1,2,0,0,0,0,1,1,1}+\frac{s\, g_{2,0,1,0,2,0,1,0,0}}{2 \epsilon -1} + \nonumber\\& + \epsilon\, \left(g_{1,1,1,-1,1,-1,1,1,1}-2 g_{1,1,0,0,0,0,1,1,1}+s\, g_{1,1,1,-1,1,0,1,1,1}-t\, g_{1,1,1,0,1,0,1,1,0}\right)\big)\Big] \nonumber
\end{align}
where the integrals are all normalized with a common overall factor $(m^2)^{2\epsilon}$ in order to guarantee the correct dimensions in dimensional regularization.
The integral topologies are depicted in  figure \ref{fig:MI}.
We performed the relevant reductions via IBP identities  \cite{Tkachov:1981wb,Chetyrkin:1981qh,Laporta:1996mq,Laporta:2001dd} by use of \texttt{FIRE} \cite{Smirnov:2008iw,Smirnov:2013dia,Smirnov:2014hma} and \texttt{LiteRed} \cite{Lee:2012cn,Lee:2013mka}.

\section{The alphabet}

After performing the change of variables \eqref{eq:changeofvars}, the differential equations for the system can be written as a $d\log$ form 
\begin{equation}\label{eq:dlog}
d f = \epsilon\, d A\, f \qquad,\qquad A = \sum_{i}\, M_i\,  \log (\text{letter})
\end{equation}
where the letters are chosen from an alphabet that includes the set of \cite{Henn:2013woa}
\begin{align}\label{eq:letters}
\text{letter}\in & \bigg\{ x, 1\pm x, y, 1\pm y, x+y, 1+ x y, x+y-4x y+x^2 y+ x y^2, \nonumber\\&\quad \frac{1+Q}{1-Q}, \frac{1+x+(1-x)Q}{1+x-(1-x)Q}, \frac{1+y+(1-y)Q}{1+y-(1-y)Q} \bigg\} \cup \dots
\end{align}
with the addition of four more letters, that we define
\begin{align}\label{eq:letters2}
\dots \cup \bigg\{ \frac{4+v+\beta }{4+v-\beta },\frac{\beta  \beta _v+4+3 v}{\beta  \beta _v-4-3 v},\frac{\beta  \beta _u+4+v \beta _u^2}{\beta  \beta _u-4-v \beta _u^2},\frac{\left((4+v) \beta _u+\beta\right) \left(\beta+(4+3 v) \beta _u\right)}{\left((4+v) \beta _u-\beta\right) \left(\beta -(4+3 v) \beta _u\right)} \bigg\}
\end{align}
We have borrowed the notation of \cite{Henn:2013woa} and \cite{Caron-Huot:2014lda} as follows
\begin{equation}
Q\equiv \sqrt{\frac{(x+y)(1+x y)}{x+y-4x y+x^2 y+ x y^2}}
\end{equation}
and
\begin{eqnarray}
& \displaystyle u\equiv \frac{s}{4m^2} \qquad & \qquad v\equiv \frac{t}{4m^2}\\
& \beta_u \equiv \sqrt{1+u} \qquad & \qquad \beta_v \equiv \sqrt{1+v}\\
& \beta_{uv} \equiv \sqrt{1+u+v} \qquad & \qquad \beta \equiv \sqrt{u\, v^2 + (4+v)^2}
\end{eqnarray}
The matrices $M$ are given in electronic form in the ancillary file \texttt{results.m}. They are sparse matrices with non-vanishing elements which, as required, do not depend on $x$, $y$ nor $\epsilon$.

\section{The solution up to order 4}

The differential equations for integrals $f_{15}-f_{17}$ and $f_{23}-f_{25}$ have nontrivial dependence on the letters \eqref{eq:letters2}.
Integral $f_{15}$ develops a dependence on them at order 3; $f_{16}$, $f_{17}$, $f_{24}$ and $f_{25}$ at order 4 and $f_{23}$ at order 5.
All other integrals and those mentioned above up to those orders are expressible in terms of Goncharov polylogarithms $G$, which in some cases reduce to harmonic polylogarithms $H$. The definitions of these functions are recalled in appendix \ref{app:polylogs} for completeness.

The solution can be determined straightforwardly in a recursive manner, order by order in $\epsilon$. Given the expansion of the vector solution $f=f^{(0)}+\epsilon f^{(1)}+\epsilon^2 f^{(2)}+...$ and using \eqref{eq:dlog} we have to recursively solve the equation $df^{(n\!+\!1)}=dA f^{(n)}$. We can integrate it in two steps by starting from an arbitrary base point and integrating on a straight horizontal line in the $(x,y)$ plane taking $y$ fixed, followed by an integration on a straight vertical line taking $x$ fixed. This is why the analytical results we obtain can be written in terms of harmonic polylogarithms with argument $x$ and Goncharov polylogarithms with argument $y$ with possible $x$-dependent indexes. Since we take the starting point to be arbitrary, this procedure fixes the solution order by order up to an integration constant unknown at each order. The latter can be fixed imposing physical conditions on the analytic structure of the result (requiring that only branch cuts associated to the cuts of the integrals are present) or comparing the results to analytic evaluations of the integrals in some limit, e.g. $y\to 1$, which is regular. We have benefited from the tools of the \texttt{HPL.m} package \cite{Maitre:2005uu,Maitre:2007kp}, for taking such limits.
For instance, the result for $f_{23}$ reads up to order 4
\begin{align}
f_{23}^{(4)} &= 8 H_0(x) G_{0,0,0}(y)+16 H_1(x) G_{0,0,0}(y)+\frac{4}{3} \pi ^2 G_{-1,0}(y)-\frac{2}{3} \pi ^2 G_{0,0}(y)+ \nonumber\\& +\frac{4}{3} \pi ^2 G_{1,0}(y)+8 G_{-1,0,0,0}(y)-8 G_{0,0,-1,0}(y)-8 G_{0,0,1,0}(y)+8 G_{1,0,0,0}(y)+ \nonumber\\&+\frac{4}{3} \pi ^2 G_0(y) H_0(x)+\frac{8}{3} \pi ^2 G_0(y) H_1(x)-\frac{4 \pi ^4}{45}
\end{align}
where the integration constant has been fixed using the information that the master integral has a vanishing limit for $y\to 1$.
Similarly, the result for integral $f_{16}$, at order 3 reads
\begin{align}
f_{16}^{(3)} &= -G_0(y) H_{0,0}(x)-2 G_0(y) H_{0,1}(x)+H_0(x) G_{-\frac{1}{x},0}(y)+2 H_1(x) G_{-\frac{1}{x},0}(y)+ \nonumber\\&-H_0(x) G_{-x,0}(y)-2 H_1(x) G_{-x,0}(y)+H_{0,0}(x) G_{-\frac{1}{x}}(y)+H_{0,0}(x) G_{-x}(y)+ \nonumber\\&+2 H_{0,1}(x) G_{-\frac{1}{x}}(y)+2 H_{0,1}(x) G_{-x}(y)+2 G_{-\frac{1}{x},-1,0}(y)-G_{-\frac{1}{x},0,0}(y)-2 G_{-x,-1,0}(y)+ \nonumber\\&+G_{-x,0,0}(y)-2 H_{-1,0,0}(x)-4 H_{-1,0,1}(x)-H_{0,1,0}(x)-2 H_{0,1,1}(x)-H_{1,0,0}(x)+ \nonumber\\&-2 H_{1,0,1}(x)+\frac{5}{6} \pi ^2 G_{-\frac{1}{x}}(y)+\frac{1}{2} \pi ^2 G_{-x}(y)-\frac{2}{3} \pi ^2 G_0(y)-\frac{4}{3} \pi ^2 H_{-1}(x)+\frac{1}{3} \pi ^2 H_0(x)+ \nonumber\\&-\frac{2}{3} \pi ^2 H_1(x)-4 \zeta (3)
\end{align}
Such expressions are collected in electronic form in the ancillary file \texttt{results.m}.
They have been successfully checked against already available results in literature and numerical integration using \texttt{FIESTA} \cite{Smirnov:2008py,Smirnov:2009pb}.
In particular, they are amenable of fast and precise numerical evaluation, for instance using \texttt{GiNaC} \cite{Bauer:2000cp,Vollinga:2004sn}.

For the integrals with a dependence on the letters \eqref{eq:letters2}, one can use their $d\log$ form, immediately available from \eqref{eq:dlog} and integrate it from a base point in the plane $(x,y)$ to given values of the Mandelstam variables.
We performed various consistency checks of this against numerical evaluations. For these we found a pedestrian numerical integration with \texttt{Mathematica}'s \texttt{NIntegrate} sufficient. 
Nevertheless, it would be interesting to ascertain whether an expression in terms of Goncharov polylogarithms could be found for these integrals, which would allow for a much more efficient numerical evaluation. 

All integrals are regular in $y\to 1$ and can be expressed in terms of harmonic polylogarithms, even those depending on the letters \eqref{eq:letters2} as they all degenerate to combinations of the $\{x,1\pm x\}$ alphabet in this limit.
We provide these limits as well in the ancillary file, expressed in terms of HPL's. Some of them can be easily simplified and give rise to extremely concise answers.
For instance, integrals $f_{24}$ and $f_{25}$ reads in this limit
\begin{align}
f_{24}^{(4)} &\underset{y\to 1}{\longrightarrow} 2 \text{Li}_4 x + \frac{\log ^4 x}{24} + 2\zeta(2) \log ^2 x + 28 \zeta(4) \\
f_{25}^{(4)} &\underset{y\to 1}{\longrightarrow} \frac13\, \log ^4\left(\tfrac{s}{m^2}\right) - \zeta(2) 
\log ^2\left(\tfrac{s}{m^2}\right) + \frac{32}{3} \zeta(3) \log\left(\tfrac{s}{m^2}\right) - 7 \zeta(4)
\end{align}
where in the last result we restored the dependence on $\tfrac{s}{m^2}$ instead of $x$ for convenience.

\section{Comments}

In this note we have been able to write a differential equation in the canonical form for the set of master integrals associated to an on-shell planar doublebox Feynman integral with two internal massive propagators and massive external momenta (see Figure \ref{fig:momentum}). This integral is a building block for the NNLO computation of massive quark-quark scattering amplitudes with full dependence on the mass. While some of the subtopologies of this system had already been computed, to our knowledge, the full seven propagator doublebox is a novel result.

The particular differential equation we found for this system can be put in $d\log$ form such that the solution of each integral can be written in terms of iterated integrals of those forms. For most of the integrals, we have been able to write them up to depth four with the use of harmonic and Goncharov polylogarithms, whose analytic properties and extensions are well known and are amenable of fast numerical evaluation. 

It would be interesting to extend this analysis to the other topologies relevant for the computation of the NNLO massive quark-quark scattering (some of them also needed for Bhabha scattering with finite electron mass). Among others, doubleboxes with different massive internal propagators routing would be needed. It would be interesting to determine if a differential equation in the canonical form for those systems exists and if they could be expressed in terms of generalized polylogarithms or elliptic functions kick in.

\section*{Acknowledgements}

We thank Andi Brandhuber, Johannes Henn, Andreas von Manteuffel for very useful comments.
MB particularly thanks Joe Hayling and Rodolfo Panerai for extra CPU's.
The work of MB was supported in part by the Science and Technology Facilities Council Consolidated Grant ST/L000415/1 \emph{String theory, gauge theory \& duality}.

\vfill
\newpage

\appendix

\section{Goncharov and harmonic polylogarithms}\label{app:polylogs}

The results of this paper are mostly expressed in terms of Goncharov and harmonic polylogarithms. In this appendix we review their definition.
Goncharov polylogarithms \cite{Goncharov:1998kja} are defined recursively as follows
\begin{align}
G_{a_1,\dots a_n} (y) = \int_0^y \frac{dt}{t-a_{1}}\, G_{a_{2}, \dots ,a_{n}} (t)
\end{align}
where
\begin{align}
G_{a_1} (y) = \int_0^y \frac{dt}{t-a_{1}}  \,, \qquad a_{1} \neq 0
\end{align}
and for $a_{1}=0$, the definition reads $G_{\underbrace{0,\dots,0}_{n}}(y) = 1/n! \log^n y$.
Harmonic polylogarithms \cite{Remiddi:1999ew} $H_{a_1,a_2,\dots,a_n}(x)$, with indices $a_i \in \{1,0,-1\}$, are defined recursively as follows
\begin{equation}\label{eq:HPL}
H_{a_1,a_2,\dots,a_n}(x) = \int_0^x  f_{a_1}(t)\, H_{a_2,\dots,a_n}(t)\, d t
\end{equation}
where
\begin{align}
& f_{\pm 1}(x)=\frac{1}{1 \mp x} \qquad f_0(x)=\frac{1}{x}\nonumber\\
& H_{\pm 1}(x)= \mp \log(1\mp x)\qquad H_0 (x)= \log x 
\end{align}
and at least one of the indices $a_i$ is non-zero.
When all indices are vanishing the definition reads
\begin{equation}
H_{\underbrace{0,0,\ldots,0}_{n}}(x) = \frac{1}{n!}\log^n x
\end{equation}

\bibliographystyle{JHEP}

\bibliography{biblio}

\end{document}